\documentclass[twoside,twocolumn,9pt]{article}
\usepackage{extsizes}
\usepackage[super,sort&compress,comma]{natbib} 
\usepackage[version=3]{mhchem}
\usepackage[left=1.5cm, right=1.5cm, top=1.785cm, bottom=2.0cm]{geometry}
\usepackage{balance}
\usepackage{times,mathptmx}
\usepackage{sectsty}
\usepackage{graphicx} 
\usepackage{lastpage}
\usepackage[format=plain,justification=justified,singlelinecheck=false,font={stretch=1.125,small,sf},labelfont=bf,labelsep=space]{caption}
\usepackage{float}
\usepackage{fancyhdr}
\usepackage{fnpos}
\usepackage[english]{babel}
\addto{\captionsenglish}{%
  
}
\usepackage{array}
\usepackage{droidsans}
\usepackage{charter}
\usepackage[T1]{fontenc}
\usepackage[usenames,dvipsnames]{xcolor}
\usepackage{setspace}
\usepackage[compact]{titlesec}
\usepackage{hyperref}
\usepackage{soul}

\usepackage{epstopdf}

\definecolor{cream}{RGB}{222,217,201}

\begin{document}

\pagestyle{fancy}

\makeFNbottom
\makeatletter
\renewcommand\LARGE{\@setfontsize\LARGE{15pt}{17}}
\renewcommand\Large{\@setfontsize\Large{12pt}{14}}
\renewcommand\large{\@setfontsize\large{10pt}{12}}
\renewcommand\footnotesize{\@setfontsize\footnotesize{7pt}{10}}
\renewcommand\scriptsize{\@setfontsize\scriptsize{7pt}{7}}
\makeatother

\renewcommand{\thefootnote}{\fnsymbol{footnote}}
\renewcommand\footnoterule{\vspace*{1pt}%
\color{cream}\hrule width 3.5in height 0.4pt \color{black} \vspace*{5pt}} 
\setcounter{secnumdepth}{5}

\makeatletter 
\renewcommand\@biblabel[1]{#1}            
\renewcommand\@makefntext[1]%
{\noindent\makebox[0pt][r]{\@thefnmark\,}#1}
\makeatother 
\renewcommand{\figurename}{\small{Fig.}~}
\sectionfont{\sffamily\Large}
\subsectionfont{\normalsize}
\subsubsectionfont{\bf}
\setstretch{1.125} 
\setlength{\skip\footins}{0.8cm}
\setlength{\footnotesep}{0.25cm}
\setlength{\jot}{10pt}
\titlespacing*{\section}{0pt}{4pt}{4pt}
\titlespacing*{\subsection}{0pt}{15pt}{1pt}

\makeatletter 
\newlength{\figrulesep} 
\setlength{\figrulesep}{0.5\textfloatsep} 

\newcommand{\topfigrule}{\vspace*{-1pt}%
\noindent{\color{cream}\rule[-\figrulesep]{\columnwidth}{1.5pt}} }

\newcommand{\botfigrule}{\vspace*{-2pt}%
\noindent{\color{cream}\rule[\figrulesep]{\columnwidth}{1.5pt}} }

\newcommand{\dblfigrule}{\vspace*{-1pt}%
\noindent{\color{cream}\rule[-\figrulesep]{\textwidth}{1.5pt}} }

\makeatother

\textbf{Enhanced energy storage density by reversible domain switching in acceptor doped ferroelectrics}
~\\
\begin{singlespace}
	
	~\\
	\textit{Zhiyang Wang$^\#$, Deqing Xue$^\#$, Dezhen Xue$^*$, Yumei Zhou$^*$, Xiangdong Ding, Jun Sun}
	
	State Key Laboratory for Mechanical Behavior of Materials, Xi'an Jiaotong University, Xi'an 710049, China
\end{singlespace}

\renewcommand*\rmdefault{bch}\normalfont\upshape
\rmfamily
\section*{}
\vspace{-1cm}


\sffamily{\textbf{
Typical ferroelectrics possess a large spontaneous polarization (\textit{P$_s$}) but simultaneously a large remnant polarization (\textit{P$_r$}) as well, resulting in an inferior energy storage density.
A mechanism that can reduce the \textit{P$_r$} while maintain the \textit{P$_s$} is demanded to enhance the energy storage property of ferroelectrics.
In the present study, it is shown that after acceptor doping and aging treatment, the domain switching in ferroelectrics becomes reversible, giving rise to a pinched double hysteresis loop.
The pinched loop with a large \textit{P$_s$} and a small \textit{P$_r$} thus results in an enhanced energy storage density.
The physics behind is a defect induced internal field that provides a restoring force for the domains to switch back.
The idea is demonstrated through a time-dependent Ginzburg-Landau simulation as well as experimental measurements in BaTiO$_3$ based single crystal and ceramics.
The mechanism is general and can be applied to various ferroelectrics, especially the environment-friendly ones.}}\\


\rmfamily 


	\section{Introduction}
Dielectric capacitors, consisting of a dielectric layer between two electrodes, store and release charges and electrical energy through the application and removal of external electric field. \cite{chu334dielectric, yang2019perovskite, pan2019ultrahigh}
They can serve as a component in a rechargeable energy-storage system of high-power/pulsed-power applications, due to their high electric power output, fast charge-discharge capability, and long cycling lifetime.  \cite{huang2019ferroelectric}
However, the low energy density limits the energy storage applications of dielectric materials in the compact and efficient electric power systems. \cite{wang2015significantly, dang2013flexible}
Therefore, new mechanisms that can significantly increase the energy density of dielectric materials are in demand.

The stored energy density of dielectric material ($J$) is determined by the applied electric field ($E$) and the electric polarization ($P$) and is given by the following integration, 
\begin{equation}
J=\int_{P_r}^{P_{max}}EdP,
\label{es}
\end{equation}
where the upper limit $P_{max}$ is the polarization under the maximum electric field, and lower limit $P_r$ is the remnant polarization ($P_r$) when the electric field is zero.
Therefore, $J$ strongly depends on $P_r$, $P_{max}$, and $E$, where the maximum of $E$ is limited by the dielectric breakdown strength ($E_b$). 
The smaller $P_r$, larger $P_{max}$ and $E_b$ are favorable for electric energy storage applications. \cite{yang2019perovskite}

\begin{figure}
	\includegraphics[width=\columnwidth]{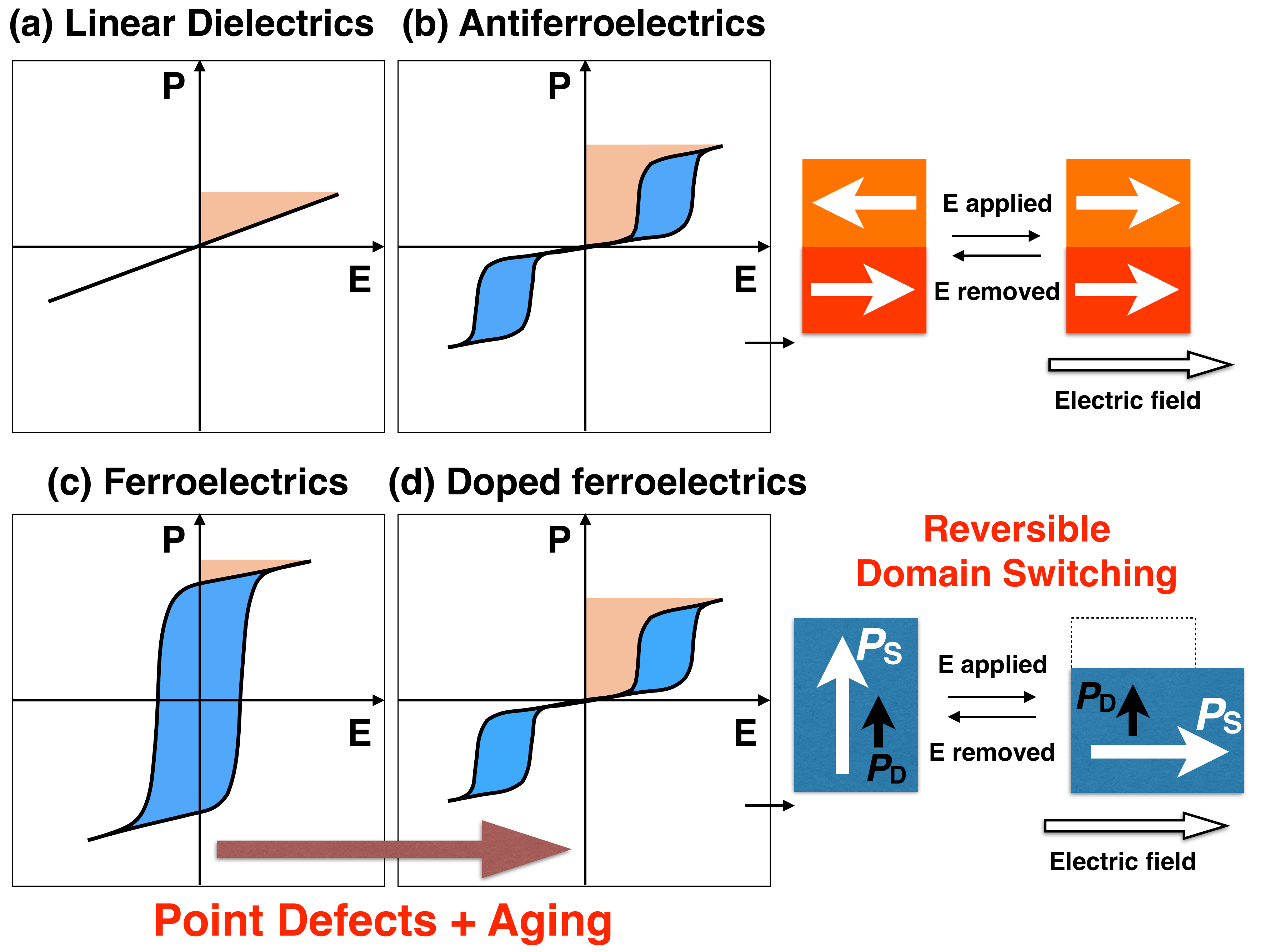}
	\centering
	\caption{Schematic illustrations of the polarization as a function of electric field for: (a) linear dielectrics; (b) ferroelectrics; (c) antiferroelectrics; (d) acceptor doped ferroelectrics after aging. The shaded areas in orange and hatched areas in blue are the recoverable energy density and dissipated energy density, respectively. The pinched double hysteresis loops with large spontaneous polarization and small remnant polarization in (c) and (d) are favorable for energy storage. The anti-parallel arrangement of dipoles in adjacent unit cell in antiferroelectrics is the ground state, which will give rise to the double hysteresis loop, as shown by the schematic beside (c). After acceptor doping and aging treatment, the domain switching in ferroelectrics becomes reversible, giving rise to a pinched double hysteresis loop, as shown by the schematic beside (d). }
	\label{4loop}
\end{figure}

Three kinds of dielectric materials, as shown in \autoref{4loop} (a), (b) and (c), are common for the production of capacitors.
A linear dielectric material responses to the applied electric field linearly according to $P = \varepsilon_0\varepsilon_r E$, which renders $P_r$ of 0, as shown in \autoref{4loop} (a). 
The corresponding $J$ is thus given by $J = \frac{1}{2}\varepsilon_0\varepsilon_r E^2$.
However, the fairly low relative dielectric constant ($\varepsilon_r$) limits its polarization and $J$. 
Thus in order to enhance $J$, many efforts have been devoted to increase the breakdown strength ($E_b$) of such materials by increasing the density, changing the architecture of devices or optimizing the microstructure. \cite{song2012improving, wang2015significantly, li2015solution, wang2017compoisitional, yin2018ultrahigh, cai2019giant}
A ferroelectric material possesses mesoscopic domains with spontaneous polarization, which can be switched along the field direction under the applied electric field.
This gives rise to a large $P_{max}$. 
However, the ferroelectric domain are unable to switch back when the electric field is removed.
Consequently, a large remnant polarization (P$_r$) occurs, as shown in \autoref{4loop} (b).
According to \autoref{es}, ferroelectric material always has a low energy-storage density, even though they have a moderate electric breakdown strength ($E_b$).
Antiferroelectric material is currently the most promising candidate for energy-storage applications. \cite{hao2014comprehensive, xu2017designing}
In antiferroelectric material, electric dipoles align in opposite directions in adjacent unit cells, leading to a zero net polarization.
Such an antiferroelectric state can be field-induced into a ferroelectric state, and thus exhibits a large net polarization (P$_{max}$) under the application of electric field. 
The anti-alignment is energetically stable and the antiferroelectric state is restored as long as the external electric field is removed, resulting a near zero P$_r$, as shown in \autoref{4loop} (c). 
Antiferroelectric materials thus perform better than ferroelectric and linear dielectric materials in energy storage, due to its double-hysteresis loop, as shown in \autoref{4loop} (c).
However, the number of antiferroelectric systems is quite limited and most of them such as La-doped Pb(Zr,Ti)O$_3$ are Pb-based ones, which causes environmental concerns. 
\cite{hao2014comprehensive}
On the contrary, the ferroelectric materials are abundant and most of them are Pb-free ones. 
Thus an alternative mechanism that can drive the ferroelectrics to have an antiferroelectric-like double hysteresis loop is needed, which will enlarge the possible candidate pool for energy storage materials, especially Pb-free ones.
In ferroelectrics, different domain states are energetically identical; thus there is no driving force to re-establish the initial multi-domain state. \cite{lines2001principles}
Consequently, a single domain state with large remnant polarization P$_r$ always appears after the removal of the field.
This inherent irreversibility in domain switching makes the potentially large energy storage of ferroelectrics futile.
Here we reported that acceptor doping and aging treatment in ferroelectrics can generate an "intrinsic" restoring force to make domain switching reversible, and consequently a small P$_r$ can be achieved without sacrificing the P$_{max}$. 
The small P$_r$ and large P$_{max}$ give rise to an enhanced energy storage property of ferroelectrics.
The acceptor dopants ({\it i.e.}, ions with valence smaller than the host ions) always generate oxygen vacancies ($V^{\cdot \cdot}_O$) in the lattice due to charge conservation. \cite{smyth2000defect}
These $V^{\cdot \cdot}_O$ are mobile and can be redistributed over a long period after sudden disturbance such as the structural phase transition, or the domain reconfiguration. \cite{ren2004large, zhang2005insitu, xue2011insitu}
In the equilibrium ferroelectric state ({\it i.e.} after aging in ferroelectric state for a long time), the polar crystal symmetry of ferroelectric phase will lead a polar distribution of $V^{\cdot \cdot}_O$. 
This is supported by the electron paramagnetic resonance (EPR) spectroscopy results, which have shown a polar alignment of the Cation-V$_O^{\cdot \cdot}$ dipoles in acceptor doped BaTiO$_3$. \cite{zhang2008reorientation, Erdem2010formation}
Such a polar alignment creates a defect polarization P$_D$ along the spontaneous polarization \textit{P$_s$} direction (P$_D$$\parallel$\textit{P$_s$}), and produces an internal bias field. \cite{ren2004large, zhou2013modeling,warren1995alignment, warren1996defect, lohkamper1990internal}
Thus within each domain of the multi-domain state, the defect polarization P$_D$ and induced internal field stabilize the spontaneous polarization $P_S$.
When such stable domains are switched by an electric field, domain switching occurs abruptly (without diffusion) with \textit{P$_s$} following the external electrical field direction. 
However, the P$_D$ cannot be rotated in such a diffusionless process, since the reorientation of P$_D$ involves the migration of $V^{\cdot \cdot}_O$. \cite{xue2009aging}
This unswitchable P$_D$ provides a restoring force or reverse internal field favoring a reverse domain switching when the electric field is removed, so a double polarization-electric field (P-E) loop is observed and the remnant polarization P$_r$ is minimized. 
Therefore, high energy storage density can be achieved.
Such a double hysteresis loop looks similar with that of the anti-ferroelectric materials, but originates from a different mechanism of reversible domain switching, as shown in \autoref{4loop} (d).
The idea is applicable to various systems, as such a phenomenon occurs in almost all the acceptor-doped ferroelectric materials.

\section{Results and Discussion}

\subsection{Modeling}

In order to verify the above idea, we first build a 2D Landau-Ginzburg model by introducing a contribution of internal field induced by defects to the Gibbs free energy of the system.
The polarization-electric field (P-E) hysteresis loop, the evolution of domain patterns under electric field as well as the energy storage property can be simulated as a function of aging time. 
Incorporating the influence of the internal field associated with defect and aging \cite{xue2012aging}, the total free-energy is written as the summation of five contributions,
\begin{equation}
G=G_{L}+G_{grad}+G_{es}+G_{em}+G_{\rho}.
\label{G1}
\end{equation}
The $G_{L}$ is the Landau expansion of the free energy in terms of the order parameter polarization ($\vec P$), which is give by,
\begin{equation}
\begin{aligned}
G_{GL}=&\int_{}^{}\alpha_1(P_x^2+P_y^2)+\alpha_{11}(P_x^4+P_y^4)+\alpha_{12}P_x^2P_y^2+\\
&\alpha_{111}(P_x^6+P_y^6)+\alpha_{112}(P_x^4P_y^2+P_x^2P_y^4)+\\
&\alpha_{1111}(P_x^8+P_y^8)+\alpha_{1112}(P_x^4P_y^4+P_x^4P_y^4)+\\
&\alpha_{1122}P_x^4P_y^4- \vec E\cdot \vec P dv,
\label{GGL}
\end{aligned}
\end{equation}
where $P_x$ and $P_y$ is the component of polarization ($\vec P$), and $\vec E$ is the external electric field applied to the ferroelectric material. 
The Ginzburg term $G_{grad}$ is the gradient energy represents the energy of the domain wall, which is given by:
\begin{equation}
G_{grad}=\int_{}^{}\frac{g_1}{2}(P_{x,x}^2+P_{y,y}^2)+\frac{g_2}{2}(P_{x,y}^2+P_{y,x}^2)+g_3P_{x,x}P_{y,y} dv,
\label{Ggrad}
\end{equation}
where $P_{i,j}$ represents the partial derivative of $P_i (i=x,y)$ with respect to $j (j=x,y)$.
The $G_{es}$ is the electrostatic energy, which represents the energy contribution of the interaction between depolarization field and dipoles.
It can be written as,
\begin{equation}
G_{es}=-\frac{1}{2}\int{\vec P}\cdot{\vec E_d}dv,
\label{Ges}
\end{equation}
where $E_d$ is the depolarization field originating from the polarization.
The $G_{em}$ is the electromechanical energy, which describes not only the pure elastic energy but also the coupling between polarization and strain. 
We define $e_1=(\epsilon_{xx}+\epsilon_{yy})/\sqrt{2}$, $e_2=(\epsilon_{xx}-\epsilon_{yy})/\sqrt{2}$ and $e_3=\epsilon_{xy}$, where the $\epsilon_{i,j}$ are the components of strain tensor.
Then the $G_{em}$ takes the form,
\begin{equation}
\begin{aligned}
G_{em}&=\int\lbrack\frac{1}{2}A_1e_1^2+\frac{1}{2}A_2e_2^2+\frac{1}{2}A_3e_3^2+\alpha e_1(P_x^2+P_y^2)\\
&+\beta e_2(P_x^2-P_y^2)+\gamma e_3P_xP_y\rbrack dv
\label{Gem}
\end{aligned}
\end{equation}
The above terms are the traditional Ginzburg-Landau free energy. 
The $G_\rho$ is the contribution associated with ferroelectric aging, which we assume to be a double-well potential similar with that of the order parameter $\vec{P}$.
The reason is that the point defect distribution symmetry follows the symmetry of the ferroelectric phase and consequently the defect polarization P$_D$ aligns along the direction of the order parameter $\vec{P}$. \cite{ren2004large, zhang2005insitu, xue2009aging, xue2011insitu}
The $G_\rho$ is given by,
\begin{equation}
\begin{aligned}
G_\rho&=\int_{}^{}\omega_1(\rho_x^2+\rho_y^2)+\omega_2(\rho_x^4+\rho_y^4)+\omega_3(\rho_x^2\rho_y^2)-\mu\vec \rho\cdot \vec{P} dv, 
\label{Grho}
\end{aligned}
\end{equation}
where $\rho_x$ and $\rho_y$ are the components of internal field vector $\vec \rho$, $\omega_i$ are the coefficients that control the aging process, and $\mu$ is the parameter that describes the coupling strength of polarization and the internal field.
It is noted that the 6th order and 8th order terms in $G_{L}$ are absent in the $G_\rho$. 
The former aims at describing the possible ferroelectric-ferroelectric phase transitions, such as the tetragonal to orthorhombic,  orthorhombic to rhombohedral in barium titanate (BaTiO$_3$). 
While only the aging behavior in cubic and tetragonal phase is considered in the present work, the 2th order and 4th order terms are  sufficient to describe the defects induced internal field. 
For more complicated behaviors of defect dipoles in orthorhombic and rhombohedral phases, the 6th and 8th order terms can be considered in the $G_\rho$.
%
%
%

%

%
The kinetics of the domain evolution are described by
the following equation,
\begin{equation}
\frac{\partial P_i}{\partial t}=-\Gamma\frac{\delta G}{\delta P_i},
\label{evolve1}
\end{equation}
where $\Gamma$ is the constant describing the evolving rate.
By solving \autoref{evolve1}, we are able to get the stable $P_x$ and $P_y$.
The distribution of $P_x$ and $P_y$ gives rise to the domain pattern of the ferroelectric system.
The evolution of the internal field is governed by the following equation,
\begin{equation}
\frac{\partial \rho_i}{\partial t}=-M\frac{\delta G_\rho}{\delta \rho_i},
\label{evolve2}
\end{equation}
By solving \autoref{evolve2}, the values of internal field at different aging time can be obtained.
We thus can calculate the polarization and domain evolution as a function of external electric field at different aging time.

\begin{table}
	\small
	\caption{Coefficients of the Ginzburg-Landau free energy and electromechanical energy from \cite{li2005phenomenological,marcel2010effects} (in SI units and \textit{T} in
		K).}
	\label{para}
	\begin{tabular*}{0.48\textwidth}{@{\extracolsep{\fill}}lll}
		\hline
		coefficient & values & unit \\
		\hline
		$\alpha_{1}$ & $4.124\times10^5(T-388)$ & $\rm{C^{-2}m^2N}$ \\
		$\alpha_{11}$ & $4.554\times10^8$ & $\rm{C^{-4}m^6N}$ \\
		$\alpha_{12}$ & $8.676\times10^8$ & $\rm{C^{-4}m^6N}$ \\
		$\alpha_{111}$ & $1.294\times10^9$ & $\rm{C^{-6}m^{10}N}$ \\
		$\alpha_{112}$ & $-1.950\times10^9$ & $\rm{C^{-6}m^{10}N}$ \\
		$\alpha_{1111}$ & $3.863\times10^{10}$ & $\rm{C^{-8}m^{14}N}$ \\
		$\alpha_{1112}$ & $2.529\times10^{10}$ & $\rm{C^{-8}m^{14}N}$ \\
		$g_{1}$ & $5\times10^{-10}$ & $\rm{C^{-2}m^{4}N}$ \\
		$g_{2}$ & $2.7\times10^{-11}$ & $\rm{C^{-2}m^{4}N}$ \\
		$g_{3}$ & $0$ & $\rm{C^{-2}m^{4}N}$ \\
		$A1$ & $2.744\times10^{11}$ & $\rm{m^{-2}N}$ \\
		$A2$ & $0.816\times10^{11}$ & $\rm{m^{-2}N}$ \\
		$A3$ & $4.88\times10^{11}$ & $\rm{m^{-2}N}$ \\
		$\alpha$ & $-1.281\times10^{10}$ & $\rm{C^{-2}m^2N}$ \\
		$\beta$ & $-0.773\times10^{10}$ & $\rm{C^{-2}m^2N}$ \\
		$\gamma$ & $-1.415\times10^{10}$ & $\rm{C^{-2}m^2N}$ \\
		\hline
		\label{tab:1}
	\end{tabular*}
\end{table}

We utilized the BaTiO$_3$ as a model system, and the coefficients of BaTiO$_3$ for the traditional Ginzburg-Landau free energy terms are listed in \autoref{tab:1}. \cite{li2005phenomenological, marcel2010effects}.
%
%
In the parameterization of $G_\rho$, we chose parameters which render the stable internal field ( \emph{i.e.}, the extremum of the $G_\rho$) larger than the coercive field of the BaTiO$_3$ of $2\times10^7 $ V/m, in order to provide sufficient restoring force to achieve reversible domain switch in acceptor doped BaTiO$_3$.
%
%
The parameters were set as $\omega_1=1\rm{V^{-2}N}$, $\omega_2=-2\omega_{1}\times 16 \times10^{14}\rm{V^{-4}m^2N}$, and $\omega_3=2\omega_2\rm{V^{-4}m^2N}$.
The coupling coefficient between the internal field $\vec \rho$ and order parameter $\vec P$, $\mu$, is set to be $1.0\times10^4$ to obtain a proper coupling strength.
The model was then built on a 128$\times$128 gird representing a 0.125$\mu$m$\times$0.125$\mu$m sample.
The two kinetic evolution equations, \autoref{evolve1} and \autoref{evolve2}, can be numerically solved simultaneously.
As the explicit form of $\frac{\delta G}{\delta P_i}$ only exits in Fourier space, \autoref{evolve1} were solved by the 3th order semi-implicit Fourier spectrum method numerically, which allows a faster and more precise solution. \cite{hu1998three}
\autoref{evolve2} were solved by the Euler method, which greatly simplifies the program but without sacrificing too much precision.
It is noted that at room temperature the domain switching has a much faster kinetic process than the diffusion of defects such as $V^{\cdot \cdot}_O$.
The time constant $\Gamma$ for domain switching is much larger than the time constant $M$ for the change of $\rho$.
Thus the internal field $\rho$ can be considered as a constant during the domain switching process.
Given that, we obtain the internal field $\rho$ by solving \autoref{evolve2} based on a stabilized domain configuration, and then include the internal field $\rho$ to the total free energy to get access to the net polarization and microstructure under external field by solving \autoref{evolve1}.
We thus can get the domain configuration and P-E hysteresis loop at different aging time ({\it i.e.,} different values of $\rho$).


%
\begin{figure}
	\centering
	\includegraphics[width = \columnwidth]{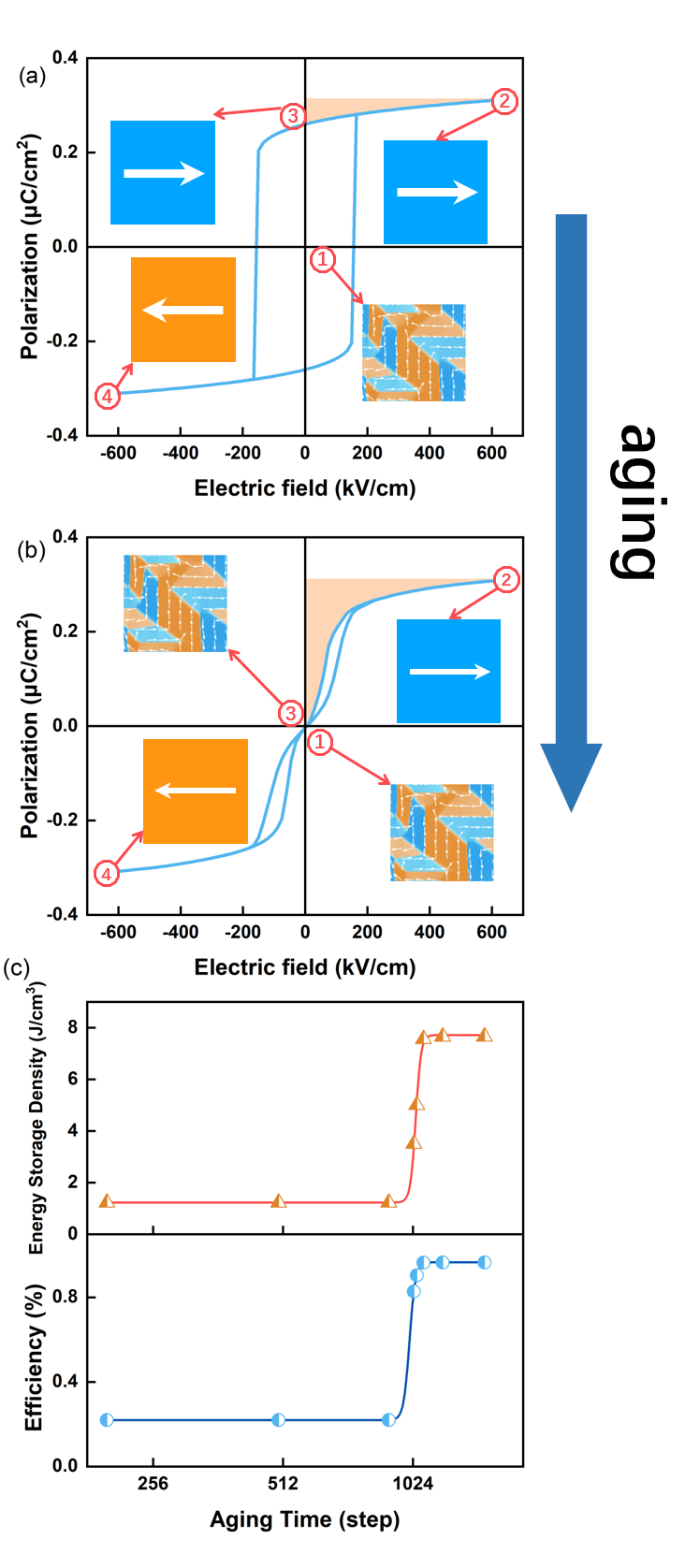}
	\caption{(a) The P-E hysteresis loops from time-dependent Ginzburg-Landau simulation for un-aged and aged ferroelectrics, respectively. After aging, a single loop turns into a double loop. The insets of (a) and (b) are typical domain patterns during loading and unloading. The initial domain pattern of aged sample in (b) can be restored after unloading, in contrast to that in (a). Both the energy storage density and the energy efficiency increase with aging time, as shown in (c). The solid line is a guide for eyes.}
	\label{fig:2}
\end{figure}

\subsection{Simulation Results}
\autoref{fig:2} shows our simulation results of defect doped BaTiO$_3$. 
We simulate the domain switching and polarization change as a function of electric field for a BaTiO$_3$ before aging ($\rho$ = 0).
As shown in \autoref{fig:2} (a), we start from the point \textcircled{1} that is a multi-domain state with zero net polarization.
With increasing electric field, the polarization increases as a result of domain switching. 
At maximum electric field, a single domain state is observed, and simultaneously polarization increases to maximum (point \textcircled{2} in the P-E curve). 
When the electric field unloads to zero, the single domain state is preserved and the polarization does not come back to zero and gives rise to the remnant polarization (point \textcircled{3} in the P-E curved).
Moreover, loading electric field in the opposite direction results in another single domain state (point \textcircled{4} in the P-E curve).
A normal single hysteresis loop is observed.
Consequently, a high remnant polarization (about 0.26 pC/cm$^2$) gives a very low energy storage density, as indicated by the red shaded part in \autoref{fig:2} (a).
We then simulate the domain switching behavior of well aged BaTiO$_3$. 
We apply electric field to a multi-domain state with net polarization of zero, as shown by point \textcircled{1} in the hysteresis loop of \autoref{fig:2}(b). 
When the field reaches maximum, a single domain configuration is observed, which corresponds to the maximum polarization (point \textcircled{2} in the P-E curve).
Interestingly, when the electric field decreases to zero, the same multi-domain pattern as the original one is recovered (compare the micrograph at points \textcircled{1} and \textcircled{3}). 
At the same time, the net polarization becomes zero.
Furthermore, the similar phenomenon occurs for the reverse electric field except that the polarization changes into negative value. 
A double P-E hysteresis loop is observed due to reversible domain-switching upon electric field cycling in the aged sample. \cite{zhang2005insitu, zhang2006aging}
The remnant polarization becomes zero, which enlarges the value of the energy storage density.
The energy storage density $J$ is then calculated from the simulated P-E curve and is plotted as a function of aging time in \autoref{fig:2} (c). 
Here we use the simulation steps when numerically solving \autoref{evolve2} to represent the aging time and the time scale is not the actual aging time.
It can be seen that when the aging time is less than 1000 steps, $J$ is small and changes a little with aging time, as the internal field $\rho$ are not able to provide enough restoring force to switch back the domains.
When the aging time is longer than 1000 steps, $J$ exhibits a sudden increase and saturates after that, which indicates that the reversible domain switching occurs.
Beside the $J$, the energy efficiency $\eta$, defined as the ratio of recoverable energy density to overall energy input density, is also calculated and is shown as function of aging time in \autoref{fig:2} (c).
As the aging induced reversible domain switching deceases the remanent polarization P$_r$, the $J$ follows the same tendency as that of $J$.
Thus aging for enough time builds an internal field by the diffusion of oxygen vacancies ($V^{\cdot \cdot}_O$), which enhances the energy storage density as well as the energy efficiency of ferroelectrics.

\subsection{Experimental Validation}

%

\begin{figure}
	\centering
	\includegraphics[width = 7.5cm]{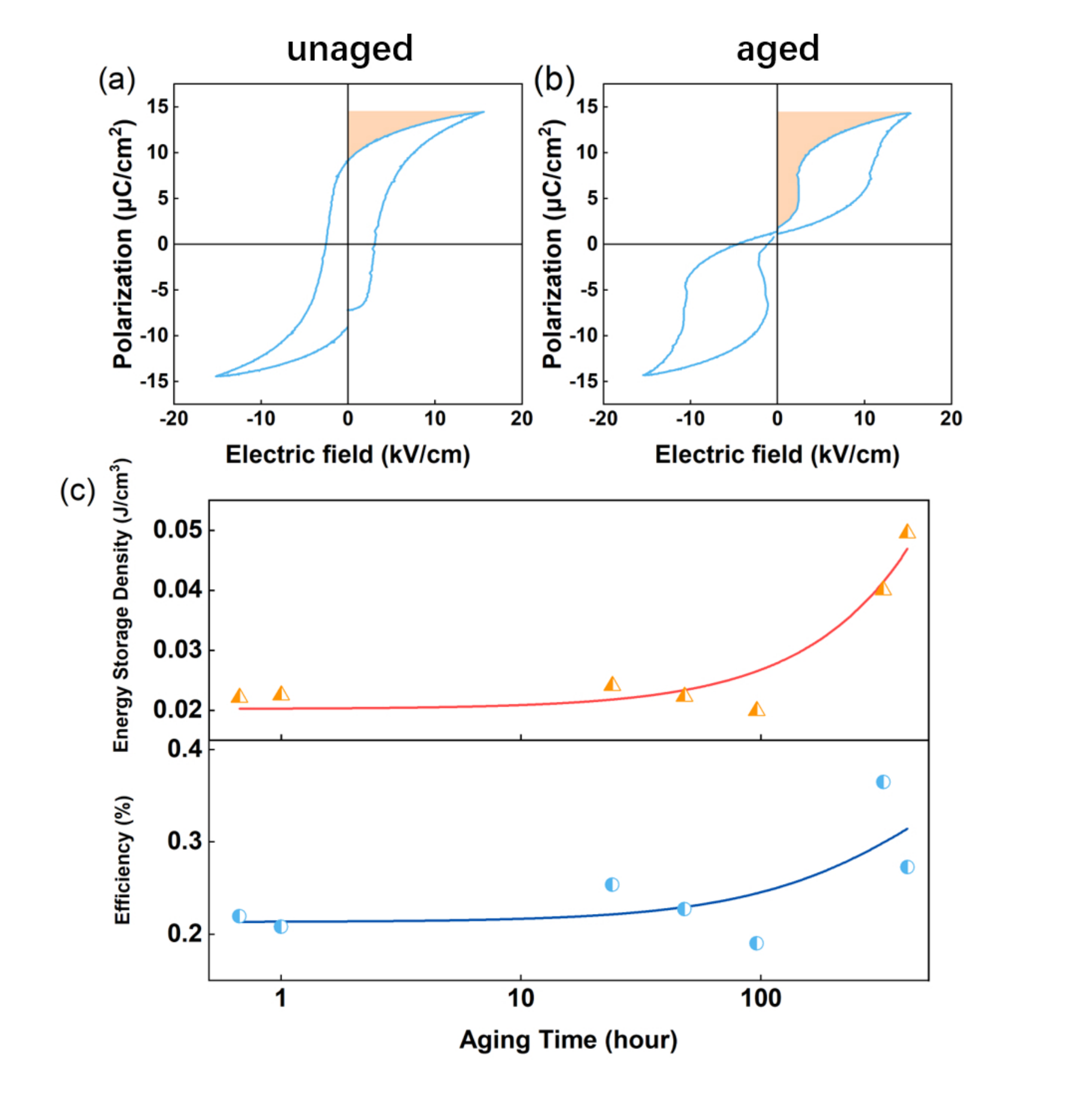}
	\caption{The P-E hysteresis loop for (a) the un-aged and (b) aged $\rm{K^+}$ doped strontium barium titanate (BST) single crystal, respectively. The shaded area of (a) and (b) are the recoverable storage energy ($J$). The aging time for (b) is 324 hours. A single loop turns into a double loop with aging time. (c) The energy storage density and energy efficiency as a function aging time. }
	The line is a guide for eyes.
	\label{sa}
\end{figure}

Experimentally, we validate our design strategy by measuring the P-E hysteresis loop of the aged and un-aged $\rm{K^+}$ doped (Ba,Sr)TiO$_3$ single crystal and $\rm{Nb^{5+}}$ and $\rm{Mn^{3+}}$ doped BaTiO$_3$ ferroelectric ceramics. 
We firstly measured the hysteresis loop of un-aged $\rm{K^+}$ doped (Ba,Sr)TiO$_3$ single crystal at room temperature.
$\rm{K^+}$ substitutes $\rm{Ba^{2+}}$ and serves as as an acceptor dopant to generate oxygen vacancies ($V^{\cdot \cdot}_O$) by charge compensation. 
A normal square P-E hysteresis loop appears in \autoref{sa} (a), because there is no defects induced internal field to induce the reversible domain switching. 
The remnant polarization is high ($9.13 \ \mu C/cm^2$) and the corresponding energy density is quite low ($0.022 J/cm^3$).
After that, we age the $\rm{K^+}$ doped (Ba,Sr)TiO$_3$ crystal at room temperature for 324 hours, and measure its P-E hysteresis loop.
As expected, a double hysteresis loop appears, as shown in \autoref{sa} (b). 
The remnant polarization $P_r$ steeply drops to $1.78 \ \mu C/cm^2$, the corresponding energy storage density is enhanced to $0.050 \times J/cm^3$, as indicated by the shadow in \autoref{sa} (b). 
Meanwhile, the energy efficiency $\eta$ also increases from 22$\%$ to 36$\%$.
%
%
To find the aging time dependence of $J$ for the single crystal sample, we then age the sample for different time and calculate $J$ by measuring the corresponding P-E loops.
In order to ensure the reliability of experiments, we de-age samples at 200 $\rm{^oC}$ for 30 minutes every time after we measure the hysteresis loop.
At 200 $\rm{^oC}$, the ferroelectric phase transforms into paraelectric phase, and the oxygen vacancies ($V^{\cdot \cdot}_O$) redistribute randomly at this temperature. \cite{xue2009aging}
Then sample is aged at room temperature for a particular time again.
The energy storage density $J$ and efficiency $\eta$ are plotted as a function aging time in \autoref{sa}(c).
Both the $J$ and $\eta$ increase with aging time. 
Such tendency is similar with the simulated result in \autoref{fig:2}(c).

\begin{figure}
	\centering
	\includegraphics[width = 6.5cm]{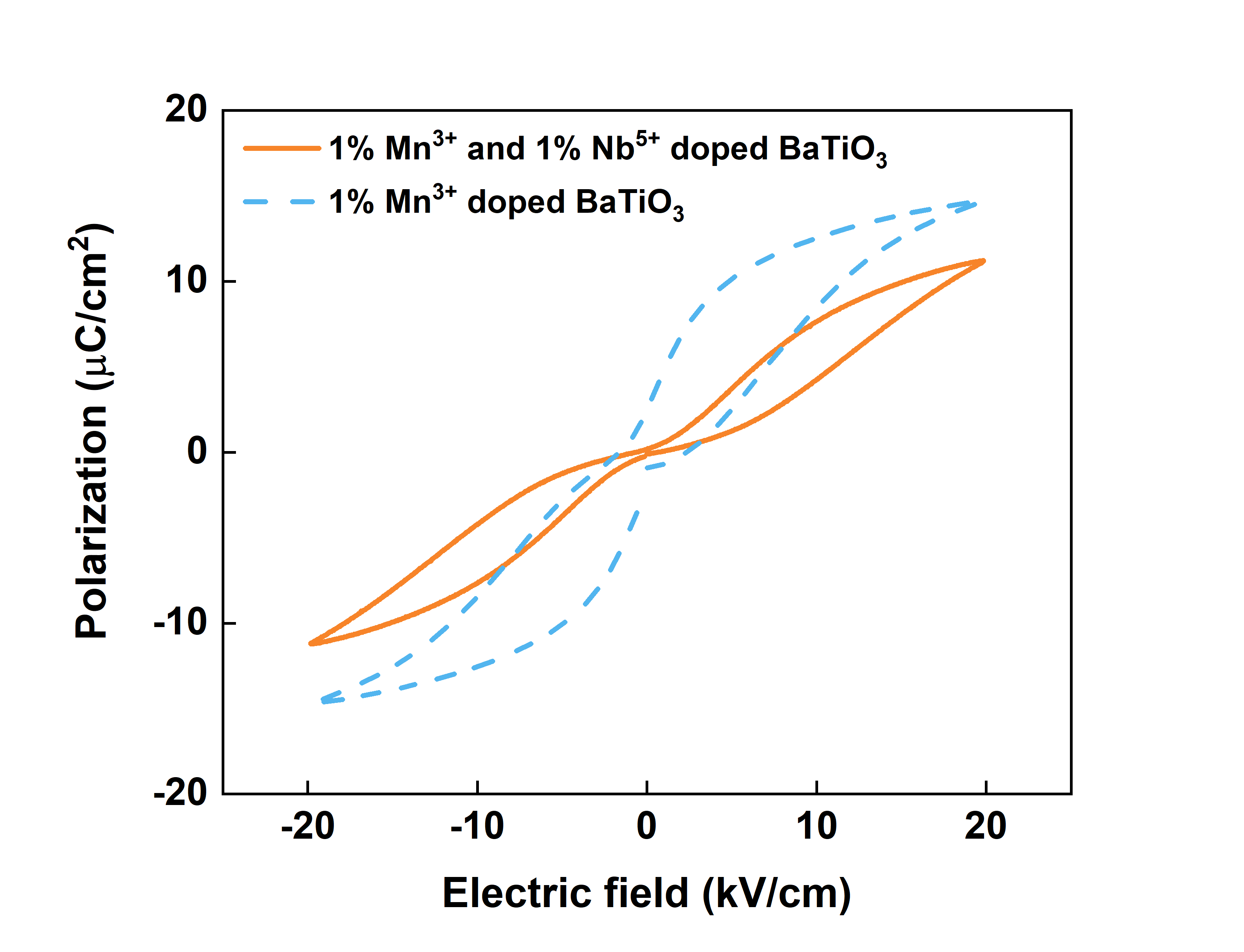}
	\caption{The P-E hysteresis loop for $\rm{Ba(Ti_{0.99}Mn_{0.01})O_{3-\delta}}$ and $\rm{Ba(Ti_{0.98}Mn_{0.01}Nb_{0.01})O_{3-\delta}}$ aging after 2700 hours.}
	\label{nb}
\end{figure}

We further demonstrate the idea in the ferroelectric ceramics, which are more widely used in industrial application.
We choose the $\rm{Nb^{5+}}$ and $\rm{Mn^{3+}}$ hybrid-doped in BaTiO$_3$ ceramics as an example.
%
%
The $\rm{Mn^{3+}}$ is doped as acceptor dopant to substitute $\rm{Ti^{4+}}$ so that oxygen vacancies ($V^{\cdot \cdot}_O$) are created by charge compensation.
Although Mn is a element with various valence states such as $\rm{Mn^{2+}}$, $\rm{Mn^{3+}}$ and $\rm{Mn^{4+}}$, the EPR data have shown that $\rm{Mn^{3+}}$ ions are dominating in Mn doped BaTiO$_3$ ceramics.  
\cite{langhammer2000crystal, kutty1985epr, lambeck1986nature, nakahara1974electronic, desu1981effect}
Adding the donor dopant $\rm{Nb^{5+}}$ increases the aging rate and decreases the coercive field and especially the energy dissipation of hysteresis.\cite{liu2006ferroelectric, bao2010control, zhou2010aging}
Thus a more obvious aging effect and an enhanced energy efficiency are expected by $\rm{Nb^{5+}}$ doping.
\autoref{nb} compares the hysteresis loops of Ba(Ti$_{0.99}$Mn$_{0.01}$)O$_{3-\delta}$ and Ba(Ti$_{0.98}$Mn$_{0.01}$Nb$_{0.01}$)O$_{3-\delta}$ after aging for 2700 hours. 
It is found that the remnant polarization P$_r$ of the sample with $\rm{Nb^{5+}}$ drops more obviously than that of BaTiO$_3$ without $\rm{Nb^{5+}}$.
It infers that $\rm{Nb^{5+}}$ and $\rm{Mn^{3+}}$ hybrid-doped BaTiO$_3$ gains a more obvious aging effects after the same aging time than $\rm{Mn^{3+}}$ doped BaTiO$_3$. 
Moreover, Ba(Ti$_{0.98}$Mn$_{0.01}$Nb$_{0.01}$)O$_{3-\delta}$ has a lower energy dissipation (\emph{i.e.}, the area between the loading and unloading curves), comparing with that of Ba(Ti$_{0.99}$Mn$_{0.01}$)O$_{3-\delta}$.
Therefore a hybrid doped BaTiO$_3$ is used in the present study.

\begin{figure}
	\centering
	\includegraphics[width = 7.5cm]{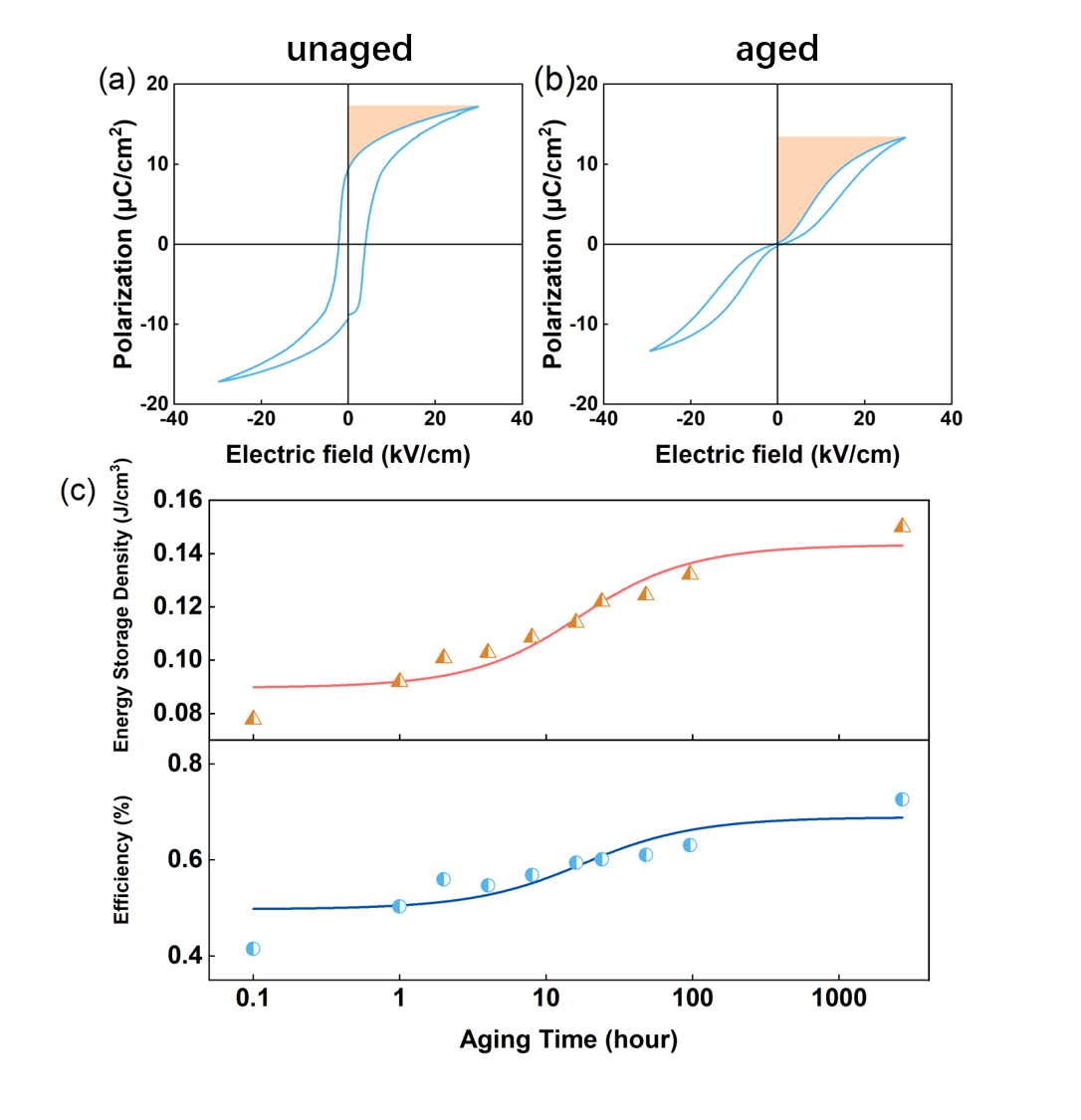}
	\caption{The P-E hysteresis loop for (a) the un-aged and (b) aged $\rm{Ba(Ti_{0.98}Mn_{0.01}Nb_{0.01})O_{3-\delta}}$ ceramic, respectively. The shaded area of (a) and (b) are the recoverable storage energy ($J$). The aging time for (b) is 2700 hours. A single loop turns into a double loop with aging time increasing. (c) The energy storage density and efficiency as a function aging time. The line is a guide for eyes.}
	\label{pa}
\end{figure}

%
The Curie temperature of the ceramic sample is about 100$\rm{^oC}$, determined by the permittivity versus temperature curve.
The experimental procedure is the same with that of the $\rm{K^{+}}$ doped (Ba,Sr)TiO$_3$ single crystal sample. 
Typical hysteresis loops of unaged and aged samples are shown in \autoref{pa}(a) and (b) respectively.
They are similar with those found in single crystal sample shown in \autoref{sa}(a) and (b). 
For the un-aged ceramic sample, a single P-E hysteresis loop, the low energy storage density of 0.077 $J/cm^3$ and efficiency of $41\%$ are observed.
On the contrary, for the aged ceramic sample, a double P-E hysteresis loop gives rise to a large energy storage density of 0.150 $J/cm^3$, about twice that of unaged sample.
Moreover, the efficiency for aged sample is increased to be 73$\%$.
The aging time dependence of the energy storage density $J$ and the efficiency $\eta$ is shown in \autoref{pa} (c).
%
%
Both $J$ and $\eta$ increase a little when the aging time is short.
After long time aging time, $J$ and $\eta$ jump suddenly, and reach a saturation value. 
The aging time dependence of $J$ and $\eta$ validates the simulation results.

\subsection{fatigue test}
\begin{figure}
	\centering
	\includegraphics[width = 7.5cm]{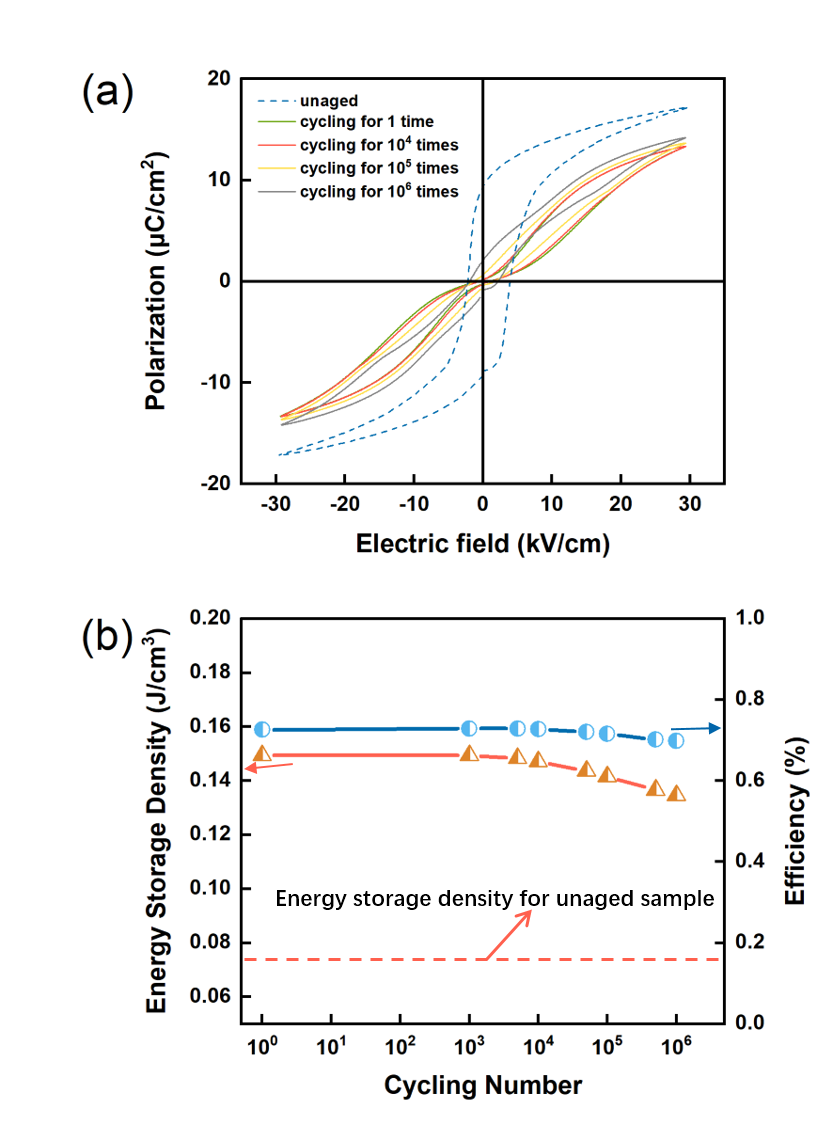}
	\caption{
	(a) The P-E hysteresis loops of $\rm{Ba(Ti_{0.98}Mn_{0.01}Nb_{0.01})O_{3-\delta}}$ ceramic after cycling for 10$^4$, 10$^5$, and 10$^6$ cycles. (b) The energy storage density and the efficiency as a function of cycling times.
	}
	\label{Fat-E}
\end{figure}

The cycling life is crucial in practical energy storage application.
We measure the fatigue performance of the aged acceptor doped ferroelectric ceramics.
An alternating electric field of 30 kV/cm is applied to a $\rm{Ba(Ti_{0.98}Mn_{0.01}Nb_{0.01})O_{3-\delta}}$ ceramic sample, which was already aged for 2700 hours. 
The P-E hysteresis loops of aged $\rm{Ba(Ti_{0.98}Mn_{0.01}Nb_{0.01})O_{3-\delta}}$ at different cycling numbers are shown in \autoref{Fat-E} (a), 
comparing with that of the unaged $\rm{Ba(Ti_{0.98}Mn_{0.01}Nb_{0.01})O_{3-\delta}}$.
%
It can be seen that the double loop with higher energy storage density persists even after cycling for 10$^6$ times.
\autoref{Fat-E}(b) further plots the energy storage density and the efficiency as a function of cycling numbers.
The energy storage density decreases with cycling. 
However, after 10$^6$ cycles, the reduction of the energy storage density is no more than 11$\%$.
It is still appropriately 80$\%$ higher than that of the unaged $\rm{Ba(Ti_{0.98}Mn_{0.01}Nb_{0.01})O_{3-\delta}}$. 
The efficiency of aged $\rm{Ba(Ti_{0.98}Mn_{0.01}Nb_{0.01})O_{3-\delta}}$ ceramic decreases a little with cycles. 
%
%
Moreover, we found that during the fatigue measurement a short suspension from electric field cycling rejuvenate the energy storage energy, which is presumably due to that a hyper-fast re-aging process leads to a recovery of energy storage density after a short rest. (See in the supplement figure 1)
Therefore, the enhanced energy storage density after aging can be potentially used for cycling applications.

\section{Conclusions}

``In our study, the obtained energy storage density of the aged BaTiO$_3$ sample is around 0.15 $J/cm^3$. 
The values of the BaTiO$_3$ based systems in literatures under the similar electric field range from 0.1 $\sim$ 0.25 $J/cm^3$.\cite{ogihara2009high, li2017novel, shen2015batio3, puli2013structure, wang2012glass, yuan2019accelerated, wang2015relaxor, hu2015dielectric}
The higher values reported are achieved either by fabricating denser and better samples to increase the dielectric break down strength or by doping ions to increase the maximum polarization and consequently the permittivity. 
Our approach of acceptor doping and aging can be further utilized to enhance the energy storage density of these reported systems \emph{via} changing the single P-E hysteresis loop to a double one. "
It should be also noted that the aging treatment may not be applicable to enhance the energy storage density of the other three kinds of dielectrics shown in \autoref{4loop}.
The double hysteresis loop of anti-ferroelectrics persists after aging but with some decay in the maximum polarization. \cite{tan2017double}
The aging affects a little to P-E hysteresis loops of linear and paraelectric materials but sometimes results in a small increase in the dielectric constant. \cite{xue2009aging}
Thus the aging treatment has a trivial influence on the energy storage density of anti-ferroelectric, linear and paraelectric materials, comparing with the ferroelectric material.

In summary, we showed that the energy storage property can be enhance through a reversible domain switching mechanism theoretically and experimentally for single crystal and polycrystalline ferroelectrics.
The acceptor doping generates the mobile oxygen vacancies while the aging treatment allows a polar distribution of these mobile defects.
Such a polar distribution induce an internal bias-field, which provides a restoring force for the domains to switch back to the initial domain states and consequently achieves a double P-E hysteresis loop with extremely low remnant polarization. 
Thus a high energy storage density, from a similar double P-E hysteresis loop with that of the anti-ferroelectrics, is realized.
Such a double hysteresis loop and the reversible domain switching can be generally achieved in many acceptor doped ferroelectric materials including BiFeO$_3$, $K_{0.5}Na_{0.5}$NbO$_3$, and Bi$_{0.5}$Na$_{0.5}$TiO$_3$-BaTiO$_3$ systems.\cite{yuan2007aging, feng2008striking, jo2010effect}
Thus the acceptor doping and aging method can be widely used in many ferroelectric systems to increase their energy storage density.
The energy dissipation and output efficiency are also important properties for ferroelectric material in energy storage application.
The hybrid-doping with acceptor and donor is an efficient way to decrease the dissipation and increase output efficiency. \cite{liu2006ferroelectric}
Studies on improving dissipation and output efficiency are expected.
The fatigue property of these acceptor doped ferroelectric material should also be considered. 
It has been shown that after more than 10$^5$ cycling, the double hysteresis loop of the aged acceptor-doped ferroelectric material maintains well. \cite{zhang2004large} 
But the upper limit of fatigue property can still be pushed forward. 
What's more, the aging time and defects doping concentration can still be optimized to obtain a higher energy storage density.

\section{Experimental Section}
The K$^+$-doped (Ba,Sr)TiO$_3$ single crystal used in the present study was grown by the KF flux method at about 1200 $^oC$. 
The as-grown samples were annealed at 1000 $^oC$ for 10 h to remove the F$^-$ so that the remaining K$^+$ was on the Ba$^{2+}$ site as an acceptor dopant and the oxygen vacancy could be created by charge compensation. 
The Ba/Sr ratio was analyzed to be about 85/15 and the concentration of K$^+$ was analyzed to be about 1.4 mol$\%$ by using the x-ray fluorescence analyzer XRF-1800 from Shimizu Corporation. 
%
%
The Curie temperature of the single crystal, determined by the permittivity versus temperature curve, is about 76 $^oC$. 
%
The samples for polarization measurement were coated with silver electrodes on both sides.

The ceramic samples were fabricated with a conventional solid-state reaction method with starting chemicals of BaCO$_3$ (99.95\%), Nb$_2$O$_5$ (99.9\%), Mn$_2$O$_3$ (99\%), and TiO2 (99.9\%).
The starting powder was ball-milled for 5 hours followed by being calcined at 1250$\rm{^oC}$ for 2 hours. 
The pre-sintered product was milled into powder again and then mixed with PVA as the glue to combine the powder together. 
The mixture was pressed into pallets under 13MPa, and then sintered at 1350$\rm{^oC}$ for 4 hours. 
%
The aging temperature was set as 80$\rm{^oC}$, which is in the ferroelectric state.
The ceramic sample for polarization measurement were coated with silver electrodes on both sides.
Ferroelectric hysteresis loops were measured with a ferroelectric tester (Radiant Workstation) at 10 Hz. 
The fatigue measurement was also conducted on the same tester at 80 Hz under 30 kV/vm alternating electric field.

\section*{Acknowledgements}
The authors gratefully acknowledge the support of National Key Research and Development Program of China (2017YFB0702401), National Natural Science Foundation of China (Grant Nos. 51671157, 51571156, and 51931004) and the 111 project 2.0 (BP2018008).

\balance
\scriptsize{
\bibliography{reference} 
\bibliographystyle{rsc} } 

\end{document}